\documentclass[conference]{IEEEtran}
\IEEEoverridecommandlockouts
\usepackage{cite}
\usepackage{amsmath,amssymb,amsfonts}
\usepackage{graphicx}
\usepackage{textcomp}
\usepackage{marvosym}
\usepackage{xcolor}
\usepackage{amsmath}
\usepackage{booktabs} 
\usepackage{mathtools}
\usepackage{algorithm}  
\usepackage{algorithmic}
\usepackage{epstopdf}
\usepackage{stfloats}
\usepackage{colortbl}
\ifCLASSOPTIONcompsoc
\usepackage[caption=false, font=normalsize, labelfont=sf, textfont=sf]{subfig}
\else
\usepackage[caption=false, font=footnotesize]{subfig}

\def\BibTeX{{\rm B\kern-.05em{\sc i\kern-.025em B}\kern-.08em
    T\kern-.1667em\lower.7ex\hbox{E}\kern-.125emX}}
\begin{document}

\title{Value of Storage for Renewable Portfolio Standard
}

\author{\IEEEauthorblockN{Jiasheng Zhang, Nan Gu, Yang Yu \textit{and} Chenye Wu$^\text{\Letter}$}
\IEEEauthorblockA{Institute for Interdisciplinary Information Sciences\\ 
Tsinghua University, Beijing, 100084, P.R. China\\
Email: chenyewu@tsinghua.edu.cn
}
}

\newtheorem{assumption}{Assumption}
\newtheorem{theorem}{Theorem}
\newtheorem{lemma}{Lemma}
\newtheorem{corollary}{Corollary}
\newtheorem{definition}{Definition}
\newtheorem{proposition}{Proposition}
\newtheorem{remark}{Remark}
\newtheorem{example}{Example}
\newcommand{\blue}[1]{{\color[rgb]{0,0,1} #1}}
\maketitle

\begin{abstract}
The ambitious targets for renewable energy penetration warrant huge flexibility in the power system. Such flexibility does not come free. In this paper, we examine the possibility of utilizing storage systems for achieving high renewable energy penetration, and identify the trade-off between providing flexibility and arbitrage against real-time prices. More precisely, we investigate the relationship among the operation cost, storage capacity, and the renewable penetration level. This illustrates the value of storage as well as the true cost induced by the high renewable penetration targets.
%    The growth of renewable brings huge indeterminacy in power system. In the meanwhile, with the amortized cost of storage decreasing to a acceptable level, this tool can help a lot to reduce such uncertainty. Plenty of researches focus on renewable problem or storage control problem solely. However, it's interesting to see what will happen with this two problems coupled. In this article, we consider a situation where an agent faces fluctuating price under a renewable quota constraint whilst she owns a storage device. The storage device can be divided into two part: one reserved for stabilizing the system and another for hedging prices. We propose an efficient algorithm to examine the value of storage, which also provides some implications.
\end{abstract}
 
\vspace{0.1cm}

\section[Introduction]{Introduction\footnote{This work has been supported in part by the Youth Program of National Natural Science Foundation of China (No. 71804087), and Turing AI Institute of Nanjing.}}
Over the past decade, many countries have set ambitious targets for renewable energy penetration, often in the (similar) form of Renewable Portfolio Standard (RPS). For example, Germany aims at a renewable energy penetration level of 80\% by the year of 2050 \cite{barbose2016retrospective}. The high requirement of RPS with only limited flexible resources leads to extremely high real time energy prices from time to time, which in turn increases the electricity bills for every household.

\subsection{Challenges and Opportunities}

% It is commonly believed that a high penetration of renewables in the power system will inevitably lead to high operation cost due to the huge uncertainties in the renewable generations. However, with the decreasing amortized cost of storage systems, the power system may embrace the urgently needed flexibility with more ease. Such flexibility, if utilized wisely, can significantly reduce the additional operation cost due to the uncertainties induced by renewables.

\begin{figure}[htbp]
        \centering  
        \includegraphics[width=7cm]{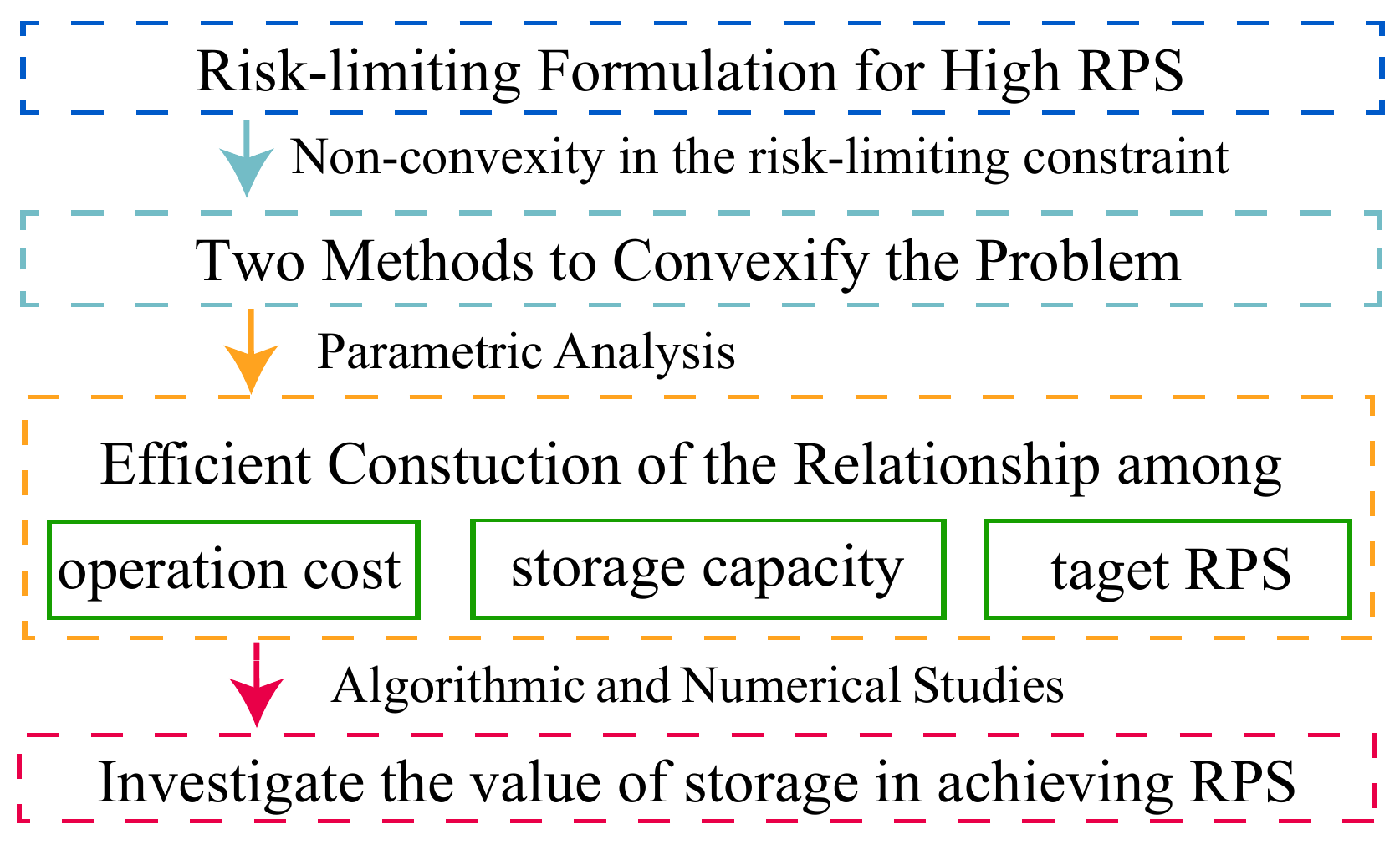}
        \caption{Paradigm to Investigate Value of Storage}%\vspace{-0.5cm}
        \label{fig 1}
    \end{figure}

While the idea is straightforward, we can utilize storage to provide more flexibility. The challenge comes from the strict requirement for RPS, which does not necessarily align with the real time price. Hence, in terms of achieving the target RPS, it is often not helpful to use storage by naively conducting arbitrage against real time price. It is more important to reserve certain capacity to meet the RPS requirement. In this paper, we seek to understand the tension between arbitraging for profits and reserving capacity for RPS. Fig. \ref{fig 1} plots the paradigm to illuminate the value of storage.%The paradigm for our research route is shown in Fig. \hyperref[fig 1]{1}.

\subsection{Related Works}
Storage is providing the vital flexibility to power system. Hence, it has been well investigated to utilize storage for high renewable energy penetration (see \cite{zhao2015review} for an excellent survey). Just to name a few, Sisternes \textit{et al.} employ the capacity expansion model of renewables to estimate the value of storage in reducing system generation cost in \cite{de2016value}. Bitar \textit{et al.} investigate the marginal value of storage system for wind power producers in \cite{bitar2019marginal}.
% The researches on renewables mostly focus on how to increase renewable penetration via economic approaches such as green certificate and feed-in tariff\cite{tamas2010feed}\cite{jensen2002interactions}. Wu \textit{et al.} proposed a algorithm to fast estimate the influence of ramping products to system generation cost since renewable penetration can cause volatility of power system in \cite{wu2015risk}. Another hotspot is storage control policy or algorithm. Wu \textit{et al.} proposed the optimal control policy for electricity storage against three-tier time of use pricing in \cite{wu2018optimal}. Qin \textit{et al.} considered two dimension of uncertainty: uncertainty in the demands and the prices then proposed a greedy algorithm to solve it in \cite{qin2015online}.
% \par Some works also take renewable penetration and price uncertainty at the same time. Chau \textit{et al.} examined the control policy for storage in the worst case when uncertain amount of renewable can be employed in \cite{chau2016cost}. Their work focus on the dynamic control while we seek to provide implications of value for storage from an algorithmic perspective. Debia \textit{et al.} estimated the marginal value of energy storage in a power market with renewable energy and thermal generation in \cite{debia2019strategic}, however, they only considered two periods since our model can be generalized to any number of stages.

% \par There are abundant related papers on the potential of storage system for renewable energy integration. \cite{castillo2014grid} and \cite{suberu2014energy} provides comprehensive reviews of energy storage techniques and insightful analysis on the advantages as well as the barriers of its application in grid-service. Specifically, from the perspective of economic benefits, \cite{de2016value} employs a capacity expansion model of renewables to estimate the benefits that energy storage can bring to reduce generation cost of the whole system. Also, \cite{bitar2019marginal} investigate how wind power producers can improve their production mechanism with the access to co-located energy storage system. 
% \par There are also researches in literature focusing on storage control policy or algorithm.
\par Another well-investigated research direction is to design the storage control policies for arbitrage. For example, van de Ven \textit{et al.} derive a threshold-based policy to minimize electricity cost for end-users facing price fluctuations in \cite{van2013optimal}. Wu \textit{et al.} propose the optimal control policy for electricity storage against three-tier time of use pricing in \cite{wu2018optimal}. Qin \textit{et al.} propose an online algorithm to address two dimension of uncertainties in demand and price in \cite{qin2015online}.
% \par This paper seeks to provide implications of value for storage under the constraints of RPS from an algorithmic perspective, which is a combination of both perspectives. 
\par Our work falls into a third group of research, which combines both perspectives.
Chau \textit{et al.} examine the control policy for storage with worst case performance guarantee in \cite{chau2016cost}. Different from their dynamic control approach, we seek to illuminate the value of storage from an algorithmic perspective. Debia \textit{et al.} estimate the marginal value of energy storage in a power market with renewable energy and thermal generation in \cite{debia2019strategic} with a focus on the two-period stylized model. In this paper, we consider a multi-stage decision making problem with risk-limiting constraints to highlight the cost induced by the high RPS. Such constraints are often non-convex, which further sophisticates the problem. We adopt the parametric functional approach \cite{wu2015risk} to address the non-convexity induced by risk-limiting constraints.
% in the worst case when uncertain amount of renewable can be employed in \cite{chau2016cost}. Their work focus on the dynamic control while we seek to provide implications of value for storage from an algorithmic perspective. Debia \textit{et al.} estimated the marginal value of energy storage in a power market with renewable energy and thermal generation in \cite{debia2019strategic}, however, they only considered two periods since our model can be generalized to any number of stages. Moreover, we underline the importance of risk-limiting constraints, which brings complexities as well as more analytical insights to the problem. To address the non-convexities, we refer to framework \cite{wu2015risk}, where Wu \textit{et al.} construct a parametric functional to optimize the risk-limiting dispatch.

\subsection{Our Contributions}
    % In this paper, we provide an algorithmic perspective of value of storage for renewable portfolio standard (RPS). The principal contributions of this paper are:
    In seek of investigating the value of storage for RPS, our principal contributions can be summarized as follows:
    \begin{itemize}
        \item \textit{Risk-limiting Formulation:} %Under proper assumptions we formulate our problem as an optimization problem and decompose it into linear part and nonlinear part.
        To evaluate the cost incurred by renewables, we use the notion of loss of load and include risk-limiting constraints in the decision making, which leads to a \emph{non-convex} optimization problem.
        
        %\item \textit{Data-driven Estimator:} %We adopt a maximum likelihood estimation (MLE) approach to fit the prediction error for renewable energy thus matching a stability requirements into the reserved capacity of storage. 
        %We adopt a maximum likelihood estimator (MLE) approach to solving the risk-limiting constraints, which provides a mapping between the reserved storage capacity and the limited risk.
        \item \textit{Problem Convexification:}
        We introduce two methods to solve the risk-limiting constraints and convexify the decision making problem. Both methods provide the mapping between the reserved storage capacity and the limited risk.
        \item \textit{Analytical Characterization:} %Instead solve the optimization problem repeatedly and blindly, we apply an efficient algorithm to construct the cost curve via storage capacity (equivalently the reservation amount). Besides, this algorithm is able to provide vivid insights for this problem.
        We design an efficient algorithm to construct the function among electricity cost, storage capacity and the target RPS. This serves as the basis for our analytical understanding of the value of storage in achieving high RPS.
    \end{itemize}
\par The rest of the paper is organized as follows.
% Section \hyperref[sec2]{\uppercase\expandafter{\romannumeral2}} formulates the problem. Section \hyperref[sec3]{\uppercase\expandafter{\romannumeral3}} introduces the MLE approach to map the system stability requirement to the desired reservation capacity of storage. In Section \hyperref[sec4]{\uppercase\expandafter{\romannumeral4}}, We start from a prototype system then formally use the algorithm for estimating the value of storage. In Section \hyperref[sec5]{\uppercase\expandafter{\romannumeral5}}, we examine our results through numerical study. We conclude this paper and discuss some extension and future works in Section \hyperref[sec6]{\uppercase\expandafter{\romannumeral6}}.
 Section \ref{sec2} introduces the system model and problem formulation. We tackle the challenges induced by the risk limiting constraints in Section \ref{sec3}. After that, Section \ref{sec4} examines the value of storage from an algorithmic perspective. We share more insights through numerical studies in Section \ref{sec5}. Finally, we deliver the concluding remarks and point out possible future directions in Section \ref{sec6}.

\section{System Model}
\label{sec2}

\begin{figure}[t]
        \centering
        \includegraphics[width=8cm]{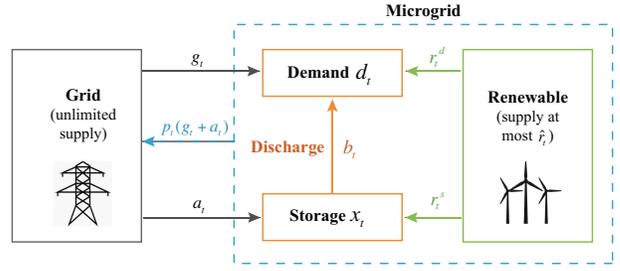}
        \caption{System Model}%\vspace{-0.5cm}
        \label{fig 2}
    \end{figure}

%We revisit a conventional storage owner model. There is an agent who invests a storage device with capacity of $B$. She is a power consumer at the same time. At each hour $t$ she decides how much quantity of electricity energy to buy from the grid with a price of $p_t$. With a horizon of $T$ hours, the storage owner faces a trade-off between power consumption and power storage. To be more specific, the agent wants to satisfy her hourly demand $d_t$ according to different situations: She can directly buy a quantity of $g_t$ from the grid to feed her demand and save $a_t$ to her storage device, or she may release $b_t$ electricity energy from the device to meet her demand, on the basis of the real-time price. We denote the accumulated power in the storage device as $x_t$. This is a typical hedge problem for storage owner. She is willing to store power at off-peak hours and release power at peak hours, facing fluctuate prices.
%\par Case is different when there is excessive renewable energy. We assume at every time there is $r_t$ \textit{free} renewable energy, which is predicted as $\hat{r_t}$. At the meanwhile, the agent's consumption is limited\ by the renewable portfolio standard (RPS), i,e, the percentage of green electricity consumed must be a constant $\alpha$, which is set in advance by the regulator. Due to the nondeterminacy of renewables, the agent preserve a space of $\Delta$ to make sure her demand can be satisfied. Fig. \hyperref[fig 2]{2} illustrates the system model.
We consider a stylized model where a microgrid operator installs certain renewables, as shown in Fig. \ref{fig 2}. To meet demand $d_t$ (predicted as $\hat{d_t}$) at time $t$, the operator could purchase energy of $g_t$ directly from grid at real time price $p_t$; utilize the renewable energy of $r_t^d$; or conduct storage control.
\par Without the requirement of RPS, the operator would simply use storage for arbitrage against the real time price for more savings in electricity bills. However, in this model, we require the operator \textit{has to} meet the target RPS of $\alpha$. This may result in the lost opportunity cost in arbitraging. Also, the system operator relies on the storage for reserving enough capacity to handle uncertainties in renewable generation.
\par To better model such uncertainties, at each time $t$, we denote the predicted and actual renewable generation by $\hat{r_t}$ and $r_t$, respectively. Facing such uncertainties and the strict RPS requirement, at each time $t$, we assume the system operator may either store energy of $a_t$ to storage, or discharge energy of $b_t$ to meet demand. In addition, the stored energy in storage may also come from the renewables, of amount $r_t^s$.
\par The microgrid system operator may seek to minimize the total electricity cost. Hence, we can formulate its decision making problem as follows:
%The storage owner problem with a constraint of RPS is modelled as follows:
\begin{alignat}{2}
        \min\quad\ & \sum\nolimits_{t=1}^T p_t(g_t+a_t)\label{eq1}\\
        s.t.\quad 
        &   g_t+b_t+r_t^d=\hat{d_t},\ \forall t, \label{eq2}\\ 
        &   \sum\nolimits_{t=1}^T (r_t^d+r_t^s)= \alpha\sum\nolimits_{t=1}^T \hat{d_t}, \label{eq3}\\
        &   x_0=x_T=0, \  x_t\ge 0,\  \forall t \label{eq4},\\
        &   x_{t+1}=x_t+a_t+r_t^s-b_t,\ t=1,...,T-1, \label{eq5}\\
        &   r_t^s+r_t^d\le \hat{r_t}, \ \forall t, \label{eq6}\\
        &   g_t\ge 0,\ a_t\ge 0,\ b_t\ge 0,\ r_t^d\ge 0, r_t^s\ge 0,\ \forall t,  \label{eq7}\\
        &   x_t\le B-\Delta,\ \forall t, \label{eq8}\\
        &   \Pr(\hat{d_t}-\hat{r_t}+\Delta \ge d_t-r_t)\ge Q\%,\ \forall t \label{eq9}
\end{alignat}
\par The decision variables are $g_t$,$a_t$,$b_t$,$r_t^d$ and $r_t^s$. We assume zero marginal cost of renewables. Hence, the cost function is only related to $g_t$ and $a_t$. Constraint (\ref{eq2}) maintains the supply demand balance, and constraint (\ref{eq3}) enforces the RPS requirement. The following two constraints are due to the evolving state-of-charge ($x_t$, at time $t$) in the storage system. Without loss of generality, we assume the decision horizon starts from mid-night when the price is often low. To achieve the maximal flexibility for the subsequent control, we require $x_0=0$. To eliminate pure arbitrage by the end of the decision making, we require $x_T=x_0$. Constraint (\ref{eq6}) implies certain amount of renewable generation may be curtailed and constraint (\ref{eq7}) indicates all the decision variables are non-negative. The final two constraints examine the value of storage. To satisfy the risk-limiting constraint $Q$, the operator may need to reserve certain storage capacity $\Delta$ as flexible resources, which in turn affects the feasible region of $x_t$'s.

\vspace{0.1cm}
\noindent \textbf{Remark}: We want to emphasize that this decision making problem is not limited to the microgrid scenario. For the grid level operation, we can directly replace the objective function with the total operation cost. Given the piece-wise linear structure of the operation cost, our subsequent analysis directly follows. We choose to use this microgrid scenario to better illustrate the notion-value of storage.

%%%%%%%%%%%%%%%%%%%%%  THIS METHOD IS INCORRECT!!!%%%%%%%%%%%%%%%%%%
% \subsection{Renewable Power Decision}
% It's obvious that the agent can achieve optimum in purchasing renewable energy as much as possible, since the renewables are attached zero cost. To reduce decision variables ($r_t^d$ and $r_t^s$), we propose a \textit{greedy} algorithm named RPQ (renewable purchase quantity) to figure out renewable purchase for demand and storage each state. This greedy algorithm returns $\hat{r_t^d}$ and $\hat{r_t^s}$, which can substitute $r_t^d$ and $r_t^s$ in the optimization problem (\hyperref[eq1]{1}) to (\hyperref[eq9]{9}) as constants.

% \begin{algorithm}[!h]
% 	\caption{RPQ$(A,p,r)$}%算法标题
% 	\begin{algorithmic}[1]%一行一个标行号
% 		\FOR{$t=1$ to $T$}
% 		\IF{$ \hat{r_t} \le \hat{d_t}$}
% 		\STATE $\hat{r_t^d}\leftarrow \hat{r_t}$;
% 		\STATE $\hat{r_t^s}\leftarrow 0$;
% 		\ELSIF{$\hat{r_t}> \hat{d_t}$}
% 		    \IF{$\hat{r_t}\le \hat{d_t}+B-\Delta-x_t$}
% 		    \STATE $\hat{r_t^d}\leftarrow \hat{d_t}$;
% 		    \STATE $\hat{r_t^s}\leftarrow \hat{r_t}-\hat{d_t}$;
% 		    \ELSE 
% 		    \STATE $\hat{r_t^d}\leftarrow \hat{d_t}$;
% 		    \STATE $\hat{r_t^s}\leftarrow B-\Delta-x_t$;
% 		    \ENDIF
% 		\ENDIF
% 		\ENDFOR
% 	\end{algorithmic}
% \end{algorithm}

%%%%%%%%%%%%%%%%%%%%%%%% SECTION 3 %%%%%%%%%%%%%%%%%%%%%%%%%%
\section{Problem Convexification}
\label{sec3}
%One of the most important decision variable is the reservation capacity $\Delta$. In order to set a proper value of $\Delta$, there is a trade-of: On the one hand, a larger reservation can warrant more stability. Namely, the probability of electricity demand to be met is greater. On the other hand, a larger $\Delta$ need to occupy more capacity which could have been used for hedging. So the capacity of storage device is divided into two parts: one part for hedge price fluctuation, the other part for enhancing the stability. 
The decision making problem is hard to solve due to the non-convexity in the risk-limiting constraint (\ref{eq9}). Note that, the parameter $Q$ only affects the reserved capacity $\Delta$ in the optimization problem. Hence, we propose to first identify the suitable $\Delta$ before decision making.

%To partial out the influence of $\Delta$, the first important observation is that the optimal value is monotone in $\Delta$.
\vspace{0.1cm}
\noindent \textbf{Fact 1}:  The total cost $\sum_{t=1}^T p_t(g_t+a_t)$ is non-decreasing in $\Delta$ in its feasible region.
\vspace{0.1cm}

This fact is based on the following observation: a larger $\Delta$ leads to a shrinked feasible region, and hence a no better total cost. This implies we can select the smallest $\Delta$ to satisfy the risk-limiting constraints: 
%\quad\textit{proof:} since the mismatch quantity is only determined by the misprediction of renewable energy, constraint (\hyperref[eq7]{7}) has no influence on other decision variables. It only imposes restriction to the feasible region of $\Delta$. Hence we just need to consider constraint (\hyperref[eq5]{5}). As $\Delta$ becomes larger, the solution domain for $x_t$ shrinks. So the optimization problem yields no-better solution since $\Delta$ increases. 
%\par Now we can pick proper a set of $\Delta$ beforehand simply according to 
\begin{align}
    \Delta(Q)\!=\!\min \{\delta\!:\!\Pr(\hat{d_t}\!-\!\hat{r_t}\!+\!\delta \ge d_t\!-\!r_t)\ge Q\%,\ \forall t\}.\label{eq10}
\end{align}
%\par Here we assume the reservation is invariant, namely, at every time slot we reserve the same quantity of storage space. However, it seems plausible that $\Delta$ can be time-varying, with respect to the demand and prediction error. We make such Homogenization assumption for the following reasons:
%\begin{itemize}
%    \item A varying reservation is hard to implement since the needs for making up mismatching are correlated among time slots. 
%    \item Prediction errors may occur beyond calculation, a global reservation concerning about the worst-case circumstance is significant.
%\end{itemize}
%\par To obtain explicit solution to $\Delta$, we need to know the probability distribution of the prediction error $e_t\coloneqq (d_t+r_t)-(\hat{d_t}+\hat{r_t})$. For the rest of the article, we make the following assumption:
%\assumption The demand prediction is accurate, namely, $\hat{d_t}=d_t$ for all $t$.\\
%\par With the assumption, it becomes sufficient to estimate $r_t-\hat{r_t}$ to obtain $\Delta$. Here let $e_t=r_t-\hat{r_t}.$
\par We choose a time-invariant $\Delta$ to sharp our understanding on the value of storage via limited parameters. In practice, one can definitely choose time varying $\Delta_t$ for more insights.
\par Unfortunately, even with this simplification, the risk-limiting constraint still display non-convex structure in general. One way to solve such non-convexity is to construct the mapping between $\Delta$ and $Q$ from the empirical prediction error distribution in demand and renewable generation. In practice, the demand prediction is rather accurate compared with that for renewable generation. Hence, in the subsequent analysis, we only consider the prediction error in the renewable generation and assume perfect demand prediction, i.e., $\hat{d_t}=d_t, \forall t$.

 \begin{figure}[t]
    \centering
	  %\subfloat[]{
       \includegraphics[width=0.7\linewidth]{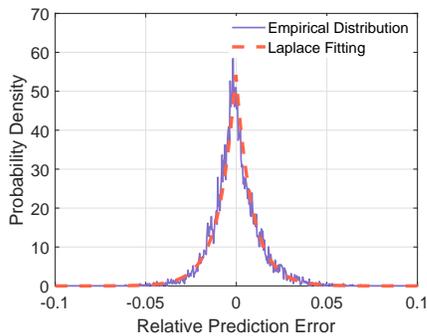}%}
%     \label{3}\hfill
% 	  \subfloat[]{
%         \includegraphics[width=0.48\linewidth]{windcdf.eps}}
%     \label{3b}
 	  \caption{Laplace Fitting for Relative Prediction Error}\vspace{-0.2cm}
	  \label{fig 3} 
\end{figure}

\par By analyzing the European  Network  of  Transmission  System Operators for Electricity (ENTSO-E) data for wind power prediction \cite{entsoe}, we observe that besides using the empirical probability distribution directly, we may also use Laplace distribution for approximation. This gives us an easier way to construct $\Delta(Q)$. Fig. \ref{fig 3} plots the fitted Laplace distribution and the relative error distribution, and Table \ref{tab 1} compares the functions $\Delta(Q)$ obtained from two approaches: $B_e$ is determined by the empirical error distribution, and $B_L$ is determined by the fitted Laplace distribution. For most cases, they are quite close. Either approach can help us convexify the optimization problem.
%\par We assume the capacity of renewables to be $10000$MW. Part of the requirement of the reservation is shown in Table \hyperref[tab 1]{1}. $B_a$ is the reservation calculated from the actual data and $B_m$ is from our modelling. 
\begin{table}[htbp]
	\centering
	\caption{Reservation $\Delta$ for Different Risk-limiting Requirements}
\begin{tabular}{cccccccc}  
		\hline  
	    $Q\%$                 & 0.7 & 0.8 & 0.9 &   0.93   &   0.96  &    0.99  &   0.998 \\
		\hline
		$B_e$(MW)   & 54.11  & 86.17 &138.3    &   162.3  &    202.4   &  322.6 & 438.9  \\
	    $B_L$(MW)  &  50.10  & 86.17 &146.3    &   174.3  &    222.4   & 342.7  &    483.0 \\
		\hline
	\end{tabular}
	\label{tab 1}
\end{table}
%%%%%%%%%%%%%%%%%%%%%%%% SECTION 4 %%%%%%%%%%%%%%%%%%%%%%%%%%
\section{Value of Storage: Algorithmic Understanding}
\label{sec4}
%    Now we turn to our core question: Given a storage capacity $B$, what's the effects of $Q$ (equivalent to $\Delta$) and $\alpha$ to the minimal cost? Of course we can simplify solve the corresponding linear programming (LP) problem (\hyperref[eq1]{1})-(\hyperref[eq8]{8}) to obtain the value in at least $O(n^2)$ complexity. However, solving the LP problem provides no implications for the value of storage and renewables. Besides, when we need multiple queries of the minimal cost, solving a LP problem many times is costly. The issue which is interesting to us is the marginal value of storage (the marginal cost of $\Delta$). We want to examine how the change of storage capacity results in change of the total cost. While the cost curve with respect to $B$ is obviously piecewise linear (since the optimization problem is linear), we are actually seeking the slope of the curve given a capacity of storage. 
 %   \par Before our formal construction of the minimal cost map, let's take a look at a toy model shown in Fig. \hyperref[fig 4]{4} at first, which can help us better understand the marginal effects of storage. 
 We seek to understand three major parameters' impacts on the minimal electricity cost: the storage capacity $B$, the risk-limiting parameter $Q$ (by the last section, it is equivalent to $\Delta$), and the RPS target $\alpha$. %Before the analytical investigation, we first use a prototype system to highlight the intuitions.

We start our analytical study by raising the following question: given a RPS target $\alpha$, how will the minimal electricity cost change with respect to the storage capacity $B$? We can define a function $C_\alpha(\beta)$ as follows to answer this question:
    \begin{align}
        \text{C}_{\alpha}(\beta)=\min  \    &  \sum\nolimits_{t=1}^T p_t(g_t+a_t) \\
                                      s.t.\ &  x_t\le \beta, \label{eq12} \\
                                             &  \alpha=\frac{\sum_{t=1}^T (r_t^d+r_t^s)}{ \sum_{t=1}^T \hat{d_t}},\\
            & \text{Constraints }(\ref{eq2}),(\ref{eq4})\text{-}(\ref{eq7}).
    \end{align}
% \par Since the problem always yields a feasible solution while $\alpha$ and $\beta$ non-negative, the domain of $\beta$ can be $[0,\infty)$. From the prototype system we have some observations as follows:
\par Using parametric LP analysis \cite{holder2010parametric}, we can show that the function $C_\alpha(\beta)$ has nice analytical properties.

\vspace{0.1cm}
\noindent \textbf{Proposition 1}: The $C_\alpha$ function is continuous, piecewise linear, convex and non-increasing in $\beta$.
\vspace{0.1cm}

We provide the detailed proof in Appendix. Here, we directly utilize the four properties to propose an efficient algorithm to construct the function $C_\alpha(\beta)$. We term the algorithm to construct $C_\alpha$ given $\alpha$ in range $[x,y]$ as Finding Breaking-points and Slopes FBS$(x,y)$, illustrated below.
% Now we can propose our algorithm \textit{Finding Breaking Points And Slopes (\hyperref[alg1]{FBS})} to construct $C_\alpha$ given $\alpha$ in $[x,y]$:
\renewcommand{\algorithmicrequire}{\textbf{Parameters:}} 
\renewcommand{\algorithmicensure}{\textbf{Output:}}
\begin{algorithm}[ht]
    \label{alg1}
    \caption{FBS($x,y$)}
	\begin{algorithmic}[1]
	    \REQUIRE The RPS target $\alpha$.
        \STATE Compute $C_\alpha(x)$ and $C_\alpha(y)$ to get their Lagrangian multipliers $\lambda_x$ and $\lambda_y$ associated with (\ref{eq12}), respectively;
        \STATE solve the system of equations below to obtain $z,c_z$;
        \begin{align}
           \left\{
                \begin{array}{rcl}
                c_z-C_\alpha(x)=\lambda_x(z-x)\\
                c_z-C_\alpha(y)=\lambda_y(z-y)
                \end{array} \right.
        \end{align}
	\IF{$c_z==C_\alpha(z)$}
	\STATE $z$ is a breaking point, the slope in $[x,z]$ is $\lambda_x$ and the slope in $[z,y]$ is $\lambda_y$;
	\STATE Return;
	\ELSE 
	\STATE Recursively construct FBS$(x,z)$ and FBS$(z,y)$;
	\ENDIF
	\end{algorithmic}
\end{algorithm}
% \par The choice of starting interval is not a big issue since we can simply select $0$ as starting point and a big value as ending point. A large value of the ending point won't have any influence to the velocity of convergence and validity of the algorithm. The constructing process for the prototype system whilst $\alpha=0.2$ is shown in Fig. \hyperref[fig 5]{5}. 

\noindent \textbf{Remark}: The initial interval $[x_0,y_0]$ for constructing $C_\alpha(\beta)$ can be determined with ease: $x_0$ can be selected as $0$ and $y_0$ can be any value that of decision maker's interests. Such arbitrary selection won't affect the algorithm efficiency as the time complexity of FBS is  $O(n)$, where $n$ is the number of breaking points in the target function. Intuitively, in the worst case, $n$ could be exponentially large in the input size of the optimization problem. Fortunately, we can show that $n$ can be polynomially bounded by the input size \cite{dey1997improved} and in practice, $n$ is fairly small even for large system\cite{wu2015risk}. %Fig. \ref{fig 5} illustrates the construction process for the prototype system when $\alpha=0.2$.
%\begin{figure}[t]
 %   \centering
%	  \subfloat[]{
 %      \includegraphics[width=0.48\linewidth]{sketch1.eps}}
 %   \label{5a}\hfill
%	  \subfloat[]{
  %      \includegraphics[width=0.48\linewidth]{sketch2.eps}}
  %  \label{5b}
%	  \caption{A Sketch for Algorithm 1 Construction.}
%	  \label{fig 5} 
%\end{figure}

%%%%%%%%%%%%%%%%%%%%%%%%%%      SECTION   5  %%%%%%%%%%%%%%%%%%%%%%%%%%
\section{Value of Storage: Numerical Studies}
\label{sec5}
\begin{figure}[htbp]
    \centering
	  \subfloat[Demand Pattern]{
       \includegraphics[width=0.48\linewidth]{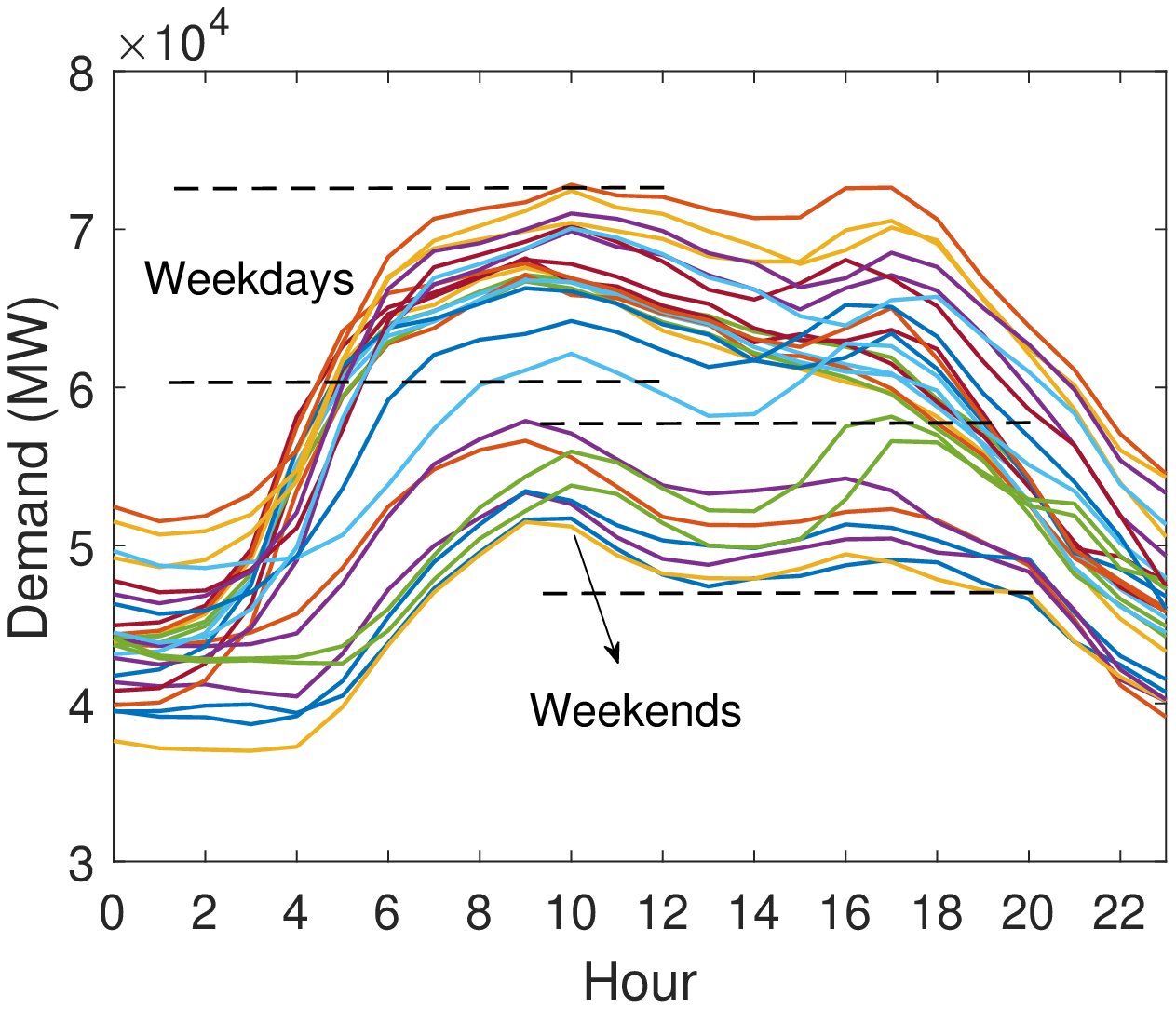}}
    \label{4(a)}\hfill
	  \subfloat[Wind Generation Pattern]{
        \includegraphics[width=0.48\linewidth]{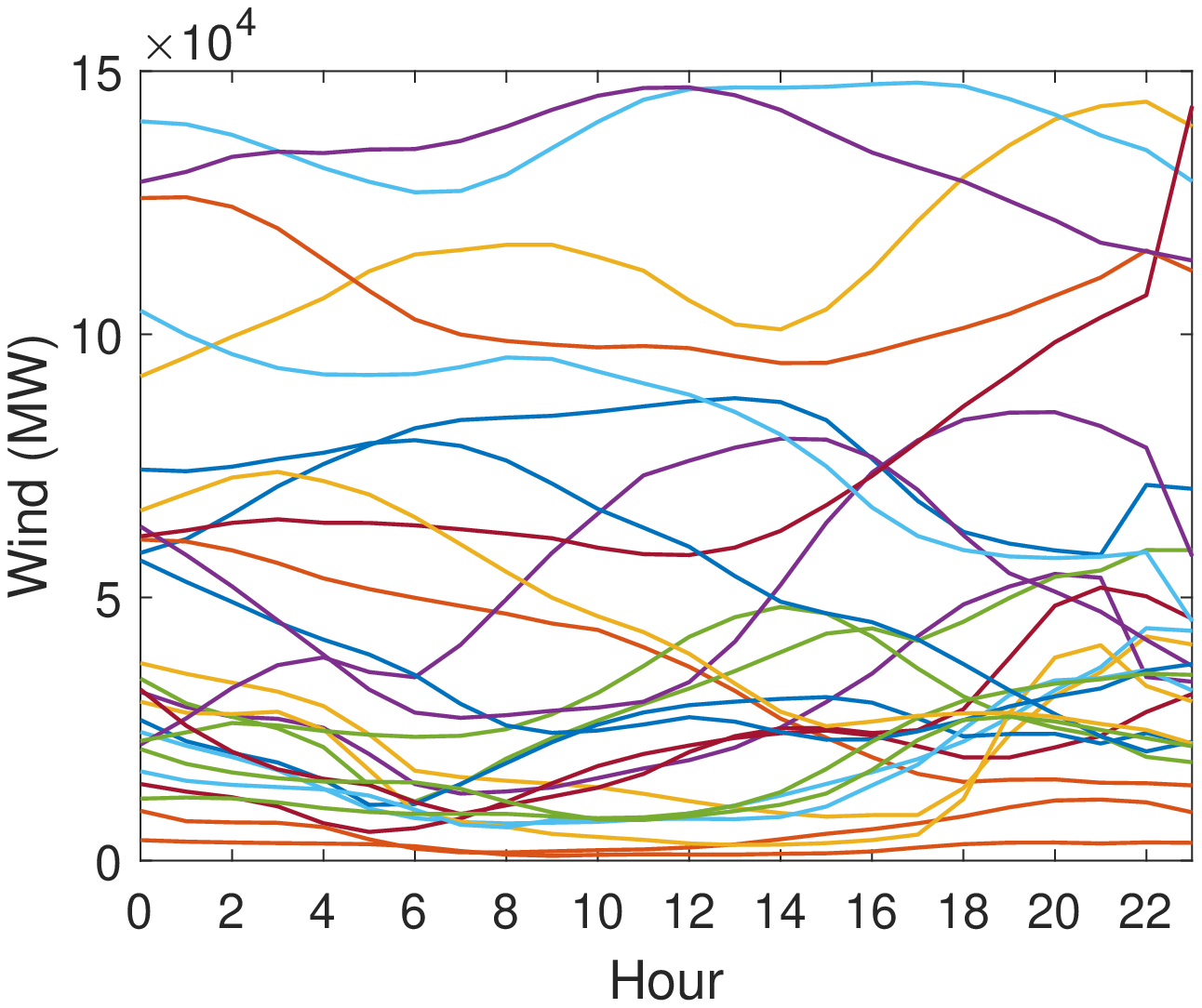}}
    \label{4(b)}
	  \caption{Sample Patterns in ENTSO-E Data of 2018-2019.}
	  \label{fig 4} 
\end{figure}

\begin{figure}[htbp]
    \centering
       \includegraphics[width=0.95\linewidth]{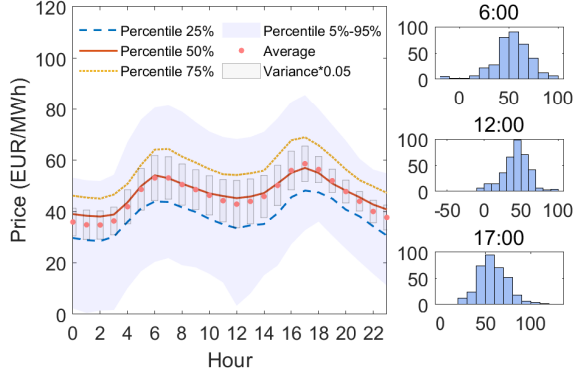}
 	  \caption{Statistical Features in Real Time Prices.}
	  \label{fig 5} 
\end{figure}

Besides the analytical properties, in this section, we want to share more insights behind the relationship function ($C_\alpha(\beta)$). We again use the ENTSO-E dataset with detailed data on annual wind generation and its forecast, annual demand and real-time price. The resolution is 1 hour. We plot some sample demand profiles and wind generation profiles in Fig. \ref{fig 4}(a) and \ref{fig 4}(b) respectively, with proper scaling. Clearly, the demand profiles display different patterns for weekdays and weekends, while wind power generation profiles highlight the stochastic nature. Fig. \ref{fig 5} visualizes the statistical features in the real time prices. From time to time, the ENSTO-E system witnesses negative prices, most likely due to strong wind. This also shows the possibility of achieving high RPS target in this system.

\begin{figure}[htbp]
    \centering
       \includegraphics[width=0.95\linewidth]{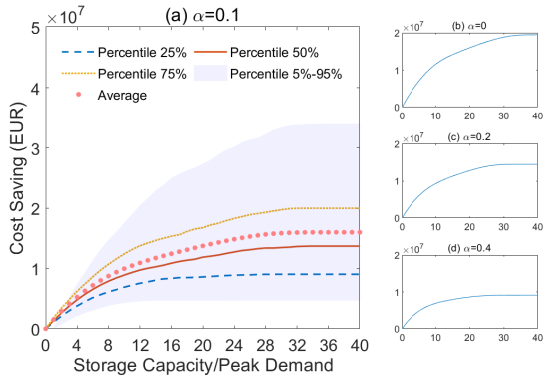}
	\caption{Cost Saving via Storage.}
    \label{fig 6} 
\end{figure}

\subsection{Cost Savings From Storage}
    The most straightforward value of storage system to the operator is cost reduction. We define cost saving 
given RPS requirement of $\alpha$ by
% We evaluate the value of storage  by observing the cost saving brought by storage investment. The \textit{cost saving} given a RPS requirement $\alpha$ is defined as:
    \begin{align}
        \text{CS}(\beta|\alpha)\coloneqq C_\alpha(\beta)-C_\alpha(0).
    \end{align}
    % The result is shown in Fig. \hyperref[fig 7]{7}. Instead of using the real capacity, we use a factor of capacity over hourly peak demand in a year as the horizontal axis. A notable conclusion from the plot is that the value (cost saving) of storage is non-increasing in $\alpha$.
    %decreases along with the requirement of renewable portfolio standard $\alpha$ increases. 
    Fig. \ref{fig 6} plots the cost saving evolves with increasing storage capacity. With different RPS targets ($\alpha=0, 0.1, 0.2, 0.4$), we conduct the simulation for a year and use different percentiles to highlight the cost saving distributions. As shown in Fig. \ref{fig 6}, even the percentile lines display certain piece-wise linear and convex structures, which verifies Proposition 1. Also, the marginal cost saving diminishes as storage capacity increases; a larger RPS requirement implies fewer cost saving: both coincides with our intuition, as dictated by Fact 1.  
    % This observation aligns with our theoretic analysis: a higher consumption of  renewable energy in the demand profile will shrink the space of storage for hedging, i.e., increase opportunity cost.

\begin{figure}[t]
    \centering 
       \includegraphics[width=0.95\linewidth]{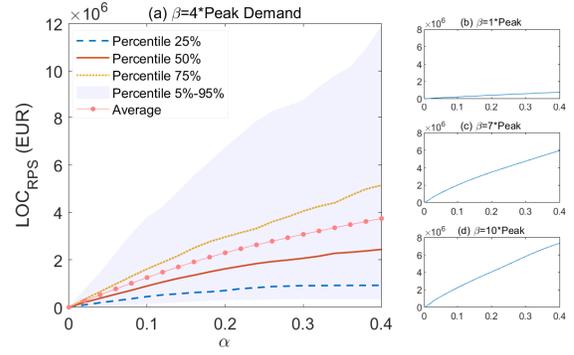}
 	  \caption{Lost Opportunity Cost for Arbitrage Caused by RPS.}
	  \label{fig 7} 
\end{figure}
\begin{figure}[t]
    \centering 
       \includegraphics[width=0.85\linewidth]{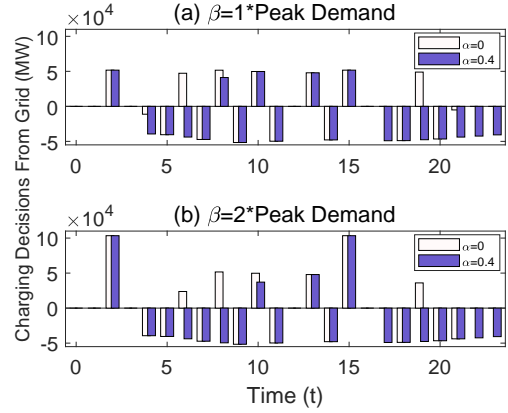}
 	  \caption{Sample Storage Control Strategies}
	  \label{fig 8} 
\end{figure}
%%%%%%%%%%%%%%   Subsection    %%%%%%%%%%%%%%%%%%%%%%
\subsection{Lost Opportunity Cost}
%  Formally, the lost opportunity for arbitrage caused by renewable penetration (LORP) under a given storage capacity $\beta$ is defined as below
We evaluate the value of storage by examining its lost opportunity cost, which is defined as follow:
\begin{align}
    \text{LOC}_\text{RPS}(\alpha|\beta)\coloneqq \text{CS}(\beta|0)-\text{CS}(\beta|\alpha).
\end{align}
Fig. \ref{fig 7} visualizes the function LOC$_\text{RPS}(\alpha|\beta)$ given different storage capacities. One might believe that larger storage capacity implies larger flexibility region for decision making, and hence leads to a lower lost opportunity cost. However, such intuition is unfounded in Fig. \ref{fig 7}. The reason is due to limited demand (more precisely, limited net demand) when RPS requirement is high, and storage cannot contribute too much in arbitrage. We highlight this counter-intuitive observation by randomly sampling a trace and plotting the optimal storage control with different parameters in Fig. \ref{fig 8}. Due to limited net demand, even when $\alpha=0$, which means the operator can simply use the renewable generation as much as possible, by doubling the battery capacity, the battery control actions do not differ too much at most time slots. This illustrates the diminishing marginal value of storage system when its capacity is large enough.
\par
\begin{figure}[t]
    \centering 
       \includegraphics[width=0.95\linewidth]{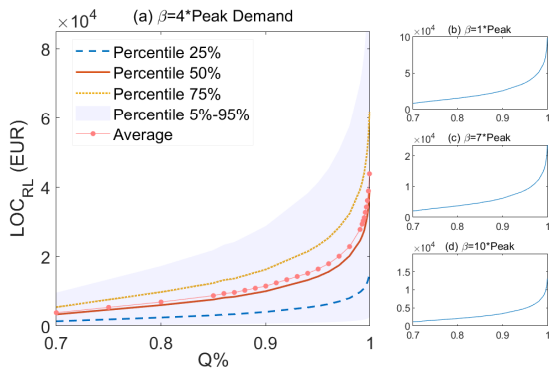}
 	  \caption{Lost Opportunity Cost of Reservation.}\vspace{-0.5cm}
	  \label{fig 9} 
\end{figure}

 On the other hand, we can define the lost opportunity cost from another perspective:
     \begin{align}
        \text{LOC}_\text{RL}(\Delta(Q)|\beta)\coloneqq C_\alpha(\beta-\Delta(Q))-C_\alpha(\beta). 
    \end{align}
This implies the lost opportunity cost by reserving capacity for the risking limiting constraints. Fig. \ref{fig 9} visualizes this function, which displays a concave feature, and as the risk limiting constraints become tight, the cost increases dramatically. This observation further emphasizes the value of storage for a stable power system.
In fact, we can use parametric analysis to show that $\text{LOC}_\text{RL}(\Delta|\beta)$ is piece-wise linear and concave in $\Delta$. 

\vspace{0.2cm}
\noindent \textbf{Remark}: We want to conclude this section by pointing out additional properties of the three functions. All the three functions enjoy double optimality. The double optimality comes from the function $C_\alpha(\beta)$, which is the minimal electricity cost given a storage capacity of $\beta$. We can prove that given certain budget of $C_\alpha(\beta)$, to satisfy the risk limiting constraints and all other constraints, the minimal required storage capacity is exactly $\beta$. This can be proved by following the route in \cite{wu2015risk}. We omit the detailed proof due to page limit.
\section{Conclusion}
\label{sec6}
In this paper, we seek to understand the value of storage system to achieve high RPS. By convexifying the decision making problem, we use parametric analysis to understand the key parameters' impacts on the decision making.
\par This work can be extended in many ways. For example, we haven't considered the network constraints in the decision making, which will help understand how the transmission line congestion affects the value of storage system. The network constraints also raise interesting discussion on the trade-off between centralized storage system and geographically distributed storage system. Also, it will be interesting to consider time varying risk limiting constraints, which will include more temporal features in the decision making.

\bibliographystyle{IEEEtran}
\bibliography{reference}

\section*{Appendix: Proof for Proposition 1}

%%%%%%%%%%%%%%%%%%%%%%%%%%%%%%%%%%%%%%%%%%%%%%%%%%%%%%%%%%%% APPENDIX
\noindent\textbf{Proof}: Continuity and piece-wise linearity are immediate results from Theorem 1.1-1.3 in \cite{holder2010parametric}. The non-increasing property has been shown in Fact 1. Hence, it suffices to show the convexity of $C_\alpha(\beta)$.
\par Let $\beta_1$ and $\beta_2$ be arbitrary realization of storage capacity $B$, such that 
\begin{align}
    \beta_2>\beta_1>0.
\end{align}
Denote the optimal solutions to $C_\alpha(\beta_1)$ and $C_\alpha(\beta_2)$ by $x_1$ and $x_2$, respectively. Hence,
\begin{align}
    x_1&=(g_1,a_1,b_1,r_1^d,r_1^s),\\
    x_2&=(g_2,a_2,b_2,r_2^d,r_2^s).
\end{align}
To prove the convexity, it suffices to examine the property of $C_\alpha(\beta)$ at $\beta^\prime=\frac{1}{2}(\beta_1+\beta_2)$. We can construct
\begin{align}
    x^\prime=\frac{1}{2}(x_1+x_2),
\end{align}
which is a feasible solution to $C_\alpha(\beta^\prime)$. Due to the property of minimization problem, we have
\begin{equation}
    \begin{aligned}
    C_\alpha\left(\frac{\beta_1+\beta_2}{2}\right)&=C_\alpha(\beta^\prime)\le\sum_{t=1}^Tp_t(g_t^\prime+a_t^\prime)\\
    &=\frac{1}{2}C_\alpha(\beta_1)+\frac{1}{2}C_\alpha(\beta_2).
\end{aligned}
\end{equation}
The convexity immediately follows. 
$\hfill\blacksquare$
% \noindent\textbf{Proof}: The continuity and piece-wise linearity comes immediately from \cite{holder2010parametric}, Theorem 1.1-1.3. The non-increasing property is due to Fact 1. What remains is the convexity. Let $\beta_1$ and $\beta_2$ be arbitrary value such that 
% \begin{align}
%     \beta_2>\beta_1>0.
% \end{align}
% Let $x^1=(g^1,a^1,b^1,r^{d1},r^{s1})$ and $x_2=(g^2,a^2,b^2,r^{d2},r^{s2})$ relatively denote the optimal solution when solving the optimization problem $C_{\alpha}(\beta_1)$ and $C_{\alpha}(\beta_2)$. For any $0\le \theta\le 1$, denote $\beta^\prime=\theta\beta_1+(1-\theta) \beta_2)$. Since the constraints for the optimization problem are all linear, it's easy to show:
% \begin{align}
%     &x^\prime=\theta x^1+(1-\theta)x^2,
% \end{align}
% also construct a feasible solution for $C_{\alpha}(\beta^\prime)$. From the property of a minimization problem we have:
% \begin{align}
%  C&_\alpha(\beta^\prime)\le  \sum\nolimits_{t=1}^T p_t(g_t^\prime+a_t^\prime)\\
%     =\theta & \sum\nolimits_{t=1}^T p_t(g_t^1+a_t^1)+(1-\theta)\sum\nolimits_{t=1}^T p_t(g_t^2+a_t^2)\\
%     =\theta & C_\alpha(\beta_1)+(1-\theta)C_\alpha(\beta_2)
% \end{align}
% So the function $C_\alpha$ is convex in $\beta$. This is consistent with the observation in the prototype system. 
%%%%%%%%%%%%%%%%%%%%%%%%%%%%%%%  APPENDIX
\end{document}